\documentclass[aps,prl,10pt,twocolumn,superscriptaddress,showpacs]{revtex4-1}
\pdfoutput=1

\usepackage[bookmarks=false]{hyperref}
\usepackage{psfrag,graphicx}
\usepackage{tikz}
\usepackage{dcolumn}
\usepackage{bm}
\usepackage{amsfonts,amssymb,amsmath} 
\usepackage{xcolor}

\usepackage{tikz}
\usetikzlibrary{decorations.markings}
\usetikzlibrary{arrows}

\usepackage{mathbbol}

\newcommand{\be}{\begin{equation}}
\newcommand{\ee}{\end{equation}}
\newcommand{\bq}{\begin{eqnarray}}
\newcommand{\eq}{\end{eqnarray}}
\newcommand{\Sp}{\,\,\,\,\,\,}

\newcommand{\rf}[1]{(\ref{#1})}

\newcommand{\tr}{\mathrm{tr}}
\newcommand{\ket}[1]{\left |#1 \right\rangle}
\newcommand{\bra}[1]{\left \langle #1 \right |}

\usetikzlibrary{positioning,shapes.misc}

\def\bra#1{\mathinner{\langle{#1}|}}
\def\ket#1{\mathinner{|{#1}\rangle}}
\let\oldto\to
\renewcommand{\to}{\!\!\oldto\!}

\begin{document}
\title{Topological Quantum Liquids with Long-Range Couplings}

\author{Kristian Patrick}
\email{py11kp@leeds.ac.uk}
\affiliation{School of Physics and Astronomy, University of Leeds, Leeds, LS2 9JT, United Kingdom}
\author{Titus Neupert}
\affiliation{University of Zurich, Department of Physics, Winterthurerstrasse 190, CH-8057 Zurich, Switzerland}
\author{Jiannis K. Pachos}
\affiliation{School of Physics and Astronomy, University of Leeds, Leeds, LS2 9JT, United Kingdom}

\date{\today}

\begin{abstract}
Very few topological systems with long-range couplings have been considered so far due to our lack of analytic approaches. Here we extend the Kitaev chain, a 1D quantum liquid, to infinite-range couplings and study its topological properties. We demonstrate that, even though topological phases are intimately linked to the notion of locality, the infinite-range couplings give rise to topological zero and non-zero energy Majorana end modes depending on the boundary conditions of the system. We show that the analytically derived properties are to a large degree stable against modifications to decaying long-range couplings. Our work opens new frontiers for topological states of matter that are relevant to current experiments where suitable interactions can be designed.

\end{abstract}

\maketitle

{\bf \em Introduction. --} 
Traditionally, physical systems are modelled by Hamiltonians that are local. This condition is linked to several quantum properties, such as the entropic area law behaviour or the existence of local order parameters. Nevertheless, models with non-local couplings become increasingly relevant in describing experiments, with Rydberg atoms~\citep{Sarma}, trapped ions~\citep{Islam} or atoms in optical cavities~\citep{YLi}, that naturally host long range interactions. Moreover, variational ans\"atze employed to model interacting systems have so far been local. While these ans\"atze have been successful in explaining important effects of interacting fermions~\citep{Bardeen,Laughlin}, our limitation in understanding the majority of strongly-correlated electron systems may be due to the intrinsic non-local nature of Coulomb interactions. 

Recognising this gap in our knowledge, an increasing number of studies have recently been focused on non-local models~\citep{Asboth, Fu, Landig, Manmana}, including Dirac~\citep{Sachdev, Cannas, Monroe, Barati, Arrachea,Porras, Fendley, Gong1, Gong2, Giampaolo} and Majorana~\citep{Degottardi, Vodola1, Vodola2, Regemortel, Delgado,Pientka1,Pientka2,Pientka3,Brydon,Rontynen} fermion chains. In these cases, non-locality is defined as there being no integer $n$ such that couplings and interactions have support over $n$ sites in the thermodynamic limit. Due to the complexity of non-local systems these studies mainly rely on numerical investigations. However, conceptual understanding often requires analytical solutions. Here we address the question, how can a system with non-local couplings be characterised topologically? Indeed, infinite-range couplings cause a meltdown of the concepts of locality and dimensionality that seem to be necessary for identifying a topological phase. Moreover, a recent study of the non-local Kitaev chain~\citep{Delgado} showed that localised end states may acquire non-zero energy, and that the topological invariant of the short-ranged model may change or completely loose its quantisation when long-range couplings are added. 

To conclusively address this problem we analytically investigate Kitaev's chain \citep{Kitaev1} decorated with infinite-range tunnelling and pairing couplings. 
%While infinite-range couplings create a fully connected graph, superconductivity imprints a one-dimensional character on the system as the pairing couplings depend on direction. 
To understand the properties of the edge modes we need to consider two distinct geometries, a semi-infinite chain and a finite chain. Via a generating function method, we analytically demonstrate that the semi-infinite chain exhibits a single localised zero energy edge mode. This zero Majorana mode is topologically stable in the same way as the edge modes of the local Kitaev chain. Moreover, its presence is in agreement with the non-trivial value of the Pfaffian topological invariant of the model, that is generalised here to open chains, thus demonstrating the persistence of topological properties in the presence of infinite-range couplings. For the physically relevant case of a finite chain the edge states from both ends hybridise due to the non-local couplings and acquire non-zero energy. The energy solution is found exactly along with the exact form of the edge modes, providing a theoretical framework for investigating localisation properties of other long-range models. Still, zero Majorana modes can be identified in the entanglement spectra encoded in the quantum correlations of an appropriately partitioned finite chain. Finally, we demonstrate the stability of our findings when the infinite-range couplings are allowed to decay. Thus, the infinite-range model shares the same topological properties as the short-range one. Note that, in the infinite range limit any attempt to close the chain results in cancellations of long-range terms. To overcome this problem we choose to restrict ourselves to an open chain, and aim to derive all of the topological properties that are present.
%Our results are relevant to current experiments with Rydberg atoms, trapped ions and atoms in optical cavities that inherently support non-local couplings.

\begin{figure}[t]
\centering
\includegraphics[width=0.8\columnwidth]{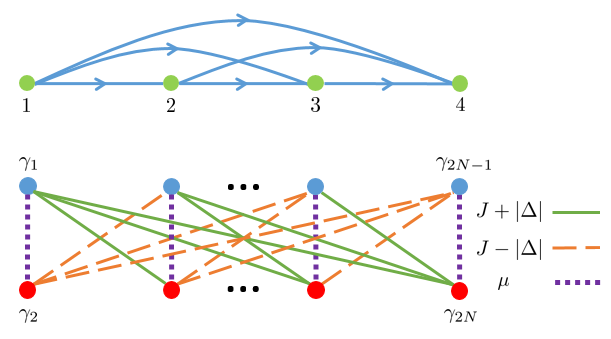}
\caption{(Top) Kitaev's chain of size $N=4$ with infinite-range couplings. While tunnelling couplings create a fully connected graph, the pairing couplings encode direction that preserves the one-dimensional character of the model.
(Bottom) The infinite-range Kitaev chain in terms of Majorana operators, $\gamma_{2j-1}$ (blue) and $\gamma_{2j}$ (red). The Majoranas are coupled with strengths $J+|\Delta|$ (green), $J-|\Delta|$ (orange) and $\mu$ (purple).
}
\label{fig:4site}
\end{figure}

{\bf \em The non-local Majorana model. --} We start from Kitaev's superconducting chain~\citep{Kitaev1} with $N$ sites, where we take the tunnelling and pairing terms to act between all pairs of fermions. The non-local superconducting Hamiltonian with open boundary conditions reads
\be
\label{eq:powerlaw}
H=\sum_{j=1}^{N}\Big[\sum_{l=1}^{N-j}\Big(\frac{J}{l^{\alpha}}a_j^{\dagger}a_{j+l}+\frac{\Delta}{l^{\alpha}} a_ja_{j+l}\Big)+ \frac{\mu}{2} a_j^{\dagger}a_j\Big] +\text{h.c.},
\ee
where $a_j$ and $a_j^\dagger$, with $j=1,...,N$, are fermionic operators, $J$ is a positive tunnelling coupling, $\Delta=|\Delta| e^{i\theta}$ is the superconducting coupling, and $\mu$ is the local chemical potential. The parameter $\alpha$ controls the range of couplings. The case $\alpha \rightarrow \infty$ corresponds to the well known local Kitaev chain. The $\alpha=0$ case corresponds to the non-decaying chain with infinite-range couplings, as shown in Fig.~\ref{fig:4site}~(Top). We initially consider the $\alpha=0$ case, before we turn to the power law behaviour with $\alpha\neq0$. 
The infinite-range model can be written, via the Jordan-Wigner transformation, as a spin chain with infinite-range cluster interactions \citep{Giampaolo,Pachos,Kay}. 

To gain an insight in the edge behaviour of the system we decompose the Dirac fermions into Majorana operators, as $a_j = e^{-i\frac{\theta}{2}}(
\gamma_{2j-1} + i \gamma_{2j})$. This transforms Hamiltonian \rf{eq:powerlaw} into a Majorana chain with infinite-range couplings, as shown in Fig.~\ref{fig:4site}~(Bottom). For the choice of couplings $\mu=0$ and $J=\pm|\Delta|$, zero energy modes emerge, localised at the end sites of the chain, much in the same way as the local Majorana Hamiltonian~\citep{Kitaev1}.

To analytically determine the energy and localisation properties of the Majorana edge states for general values of the couplings $\mu$, $J$ and $\Delta$, we need to construct a systematic method. Our approach is based on the generating function that is routinely used to probe edge characteristics of local models \citep{Pershoguba,Kon,Sticlet}. Assume the state $\ket{\psi}=\sum_{j=1}^{N}\chi_j^{\dagger}\psi_j\ket{0}$ is an eigenstate of Hamiltonian \rf{eq:powerlaw} with energy $E$, where $\chi_j^{\dagger}=\begin{pmatrix}a_j^{\dagger}&a_j\end{pmatrix}$, $\psi_j=\begin{pmatrix}\psi_j^A \\ \psi_j^B\end{pmatrix}$ is the amplitude of the state at site $j$ and $\ket{0}$ is the superconducting vacuum. The equation $H\ket{\psi}=E\ket{\psi}$ leads to the recursion relation 
\be
\label{eq:recursion}
\sum_{l=1}^{N}\Big[\Gamma_2\psi_{j+l}+\Gamma_2^{\dagger}\psi_{j-l}\Big]+(\Gamma_1-2E)\psi_j =0 ,
\ee
where $\Gamma_1$ and $\Gamma_2$ contain the coupling constants $J$, $\Delta$, and $\mu$.
%$\Gamma_1=\begin{pmatrix}
%\mu & 0 \\
%0 & -\mu
%\end{pmatrix}$ 
%and $\Gamma_2=\begin{pmatrix}
%J & -\Delta^* \\
%\Delta & -J
%\end{pmatrix}$.
By defining $\Psi_j=\sum_{k=j}^{N}\psi_k$, Eq.~\rf{eq:recursion} takes the simpler form $ M\Psi_{j}+K\Psi_{j+1}+\Gamma_2^{\dagger}\Psi_1 = 0$, where $M=\Gamma_1-\Gamma_2^{\dagger}-2E$ and $K=\Gamma_2-\Gamma_1+2E$. Multiplying this by $z^j$ for $z\in\mathbb{C}$, summing over all $j$, and defining the generating function as 
\be
G(z)=\sum\limits_{j=1}^Nz^{j-1}\Psi_j,
\label{eq:defG}
\ee
we obtain an equation for $G(z)$ that can be directly solved to give
\be
\label{eq:GF}
G(z)= (K+zM)^{-1}\left[K-\Gamma_2^{\dagger}\frac{z(1-z^N)}{1-z}\right]\Psi_1.
\ee
For concreteness, we take $\Psi_1= \begin{pmatrix}\phi_1 \\ \phi_2\end{pmatrix}$, where $\phi_1$ and $\phi_2$ are to be determined. 

From the definition of $G(z)$ given in  \rf{eq:defG} we see that the properties of its poles depend on the distribution of the amplitudes $\psi_j$ of the modes along the chain. The closed form~\rf{eq:GF} allows us to study the poles of $G(z)$. Apart from the explicit $z=1$ pole, the rest can be produced by the inverse matrix $(K+zM)^{-1}$. For that we consider the zeroes $\{z_1,z_2\}$ of $\det (K+zM)=\big[4E^2-(J-\mu)^2\big](1-z)^2+|\Delta|^2(1+z)^2$ that are not cancelled from the rest of the expression. Vieta's formula relates the zeroes of $\det (K+zM)$ with the ratio of the coefficients of $z^0$ and of $z^2$, giving $z_1z_2=1$ \citep{Vieta}. As zeroes of the determinant with $z_i=1$ correspond to bulk modes, we are interested in the zeroes $\{z_1,z_2\}$ that have amplitudes larger or smaller than one (for more details see Supplemental Material).

{\bf \em Zero energy Majorana edge modes in a semi-infinite chain. --} We first consider the case of a semi-infinite chain extending to $+\infty$ with a single boundary at $j=1$. The definition \rf{eq:defG} of the generating function gives that in the case of an edge state, $G(z)$ only diverges exponentially as $z$ approaches the singular points $z_i$ that satisfy $|z_i|>1$. Hence, all zeroes $z_i$ of $\det (K+zM)$ in the denominator of \rf{eq:GF} with $|z_i|<1$ or $|z_i|=1$ need to be cancelled, along with the explicit $z=1$ pole in \rf{eq:GF}. 
%To study this regime we restrict from now on to $|z|<1$. With this condition we can substitute $z^N\to 0$ in Eq.~\eqref{eq:GF} for $N$ large. This simplifies the generating function without losing any information about the poles. 
%The cancellation of the zeros of $\det (K+zM)$ and the $z=1$ pole can happen if both of the components of the numerator have the same zeroes. 
Applying these conditions to the generating function results, for $\mu < J$, in a Majorana edge state localised at $j=1$ with energy $E=0$ and wave function
\be
\psi_j \propto z_1^{-j}\begin{pmatrix}
e^{-i\theta}\\
1
\end{pmatrix},
\label{eq:semi}
\ee
where $\theta$ is the phase of the superconducting coupling $\Delta$ and $z_1=1-\frac{2J}{\mu}$.  

Finding zero energy modes is a manifestation of the system being in a topological phase. This is a surprising result given that the presence of the non-local couplings dramatically alters the notion of locality and dimensionality. We will later connect the presence of the zero modes with the non-trivial value of a topological invariant that characterises the topological phase of the non-local chain. 

The zero Majorana edge state~\eqref{eq:semi} of the semi-infinite chain is the unique, spectrally isolated state, which is both an eigenstate of the Hamiltonian and of the particle-hole operator ${\cal P}=\mathcal{K} \sigma_2$, where $\mathcal{K}$ is complex conjugation and $\sigma_2$ the second Pauli matrix. This property pins the Majorana edge state to $E=0$. If the system is deformed away from the exactly solvable point by a series of unitary operations (local or not) that commute with particle-hole symmetry, this  property is preserved as long as the energy gap does not close. Hence, the edge mode~\eqref{eq:semi} is topologically stable in the same way as the edge modes of the local Kitaev chain~\citep{Kitaev1}.

{\bf \em Gapped edge modes in a finite chain. --} We now turn to the case of a finite chain of $N$ sites with two boundaries, as shown in Fig.~\ref{fig:4site}~(Top). From this fully connected graph it is possible to distinguish its two ends through the pairing terms of Hamiltonian~\rf{eq:powerlaw}, that naturally encodes a direction. This is apparent when we rewrite the Hamiltonian using Majorana operators, as shown in Fig.~\ref{fig:4site}~(Bottom). In addition to the exponential singularities at $z_i$ for $|z_i|>1$ that we met in the semi-infinite case, the edge state generating function of a finite chain also diverges linearly with the number of sites $N$ at $z=1$ (see Supplemental Material). 

To proceed, we first consider the symmetries of Hamiltonian \rf{eq:powerlaw} with two boundaries that are absent on a semi-infinite chain. These symmetries, present also in a local chain, impose additional conditions that are useful to analytically determine the non-zero energy of the edge modes. We observe that a reflection of the chain about its centre, $j\to N-j+1$ with $l\to -l$, as well as a substitution of $a_j\to -a_j$, $a_j^{\dagger}\to a_j^{\dagger}$ transform the Hamiltonian as $H\to -H$, that give $\psi_j^A=-\psi_{N-j+1}^A$ and $\psi_j^B=\psi_{N-j+1}^B$. A second set of possible transformations, including a reflection along with $a_j\to a_j$ and $a_j^{\dagger}\to -a_j^{\dagger}$, that also transform the Hamiltonian as $H\to -H$, give $\psi_j^A=\psi_{N-j+1}^A$ and $\psi_j^B=-\psi_{N-j+1}^B$. These conditions impose either $\phi_1=\sum_j^N\psi_j^A=0$ or $\phi_2=\sum_j^N\psi_j^B=0$, respectively. 

We first set $\phi_2=0$ in Eq.~\eqref{eq:GF}. As we have seen above, demanding that the generating function describes an exponentially localised mode imposes conditions on its divergence properties. These conditions determine that, for $\mu<J$, the energy of the mode is given by
\be
\label{eq:energy}
E_+=\frac{J-\mu}{2}\hspace{2mm}\frac{J^2-|\Delta|^2}{J^2+|\Delta|^2},
\ee
and the amplitudes of the corresponding state are given by
\be
\psi^{+}_j \propto z_1^{-j}\begin{pmatrix}
1\\
\frac{J}{\Delta^{*}}
\end{pmatrix}+ z_1^{j-N-1}\begin{pmatrix}
1\\
-\frac{J}{\Delta^*}
\end{pmatrix},
\label{eq:edge1}
\ee
where $z_1=\frac{2E_++2J-\mu}{2E_+-\mu}$ is the inverse of the cancelled pole. When we set $\phi_1=0$, we obtain the second energy, $E_-=-E_+$, with the mode given by $\psi^{-}_j\propto\left[\sigma_1\psi^{+}_j\right]^*$.
%\be
%\psi^{-}_j \propto z_1^{-j}\begin{pmatrix}
%\frac{J}{\Delta}\\
%1
%\end{pmatrix}+ z_1^{j-N-1}\begin{pmatrix}
%-\frac{J}{\Delta}\\
%1
%\end{pmatrix}.
%\label{eq:edge2}
%\ee
Hence, the localised Majorana edge modes of a finite chain have in general non-zero energies $E_\pm$ determined by Eq.~\rf{eq:energy} that are independent of the system size $N$. 

The non-zero value of the energies $E_\pm$ is due to the non-local coupling between the edge modes at the two ends regardless of their distance. These couplings hybridise the two Majorana edge modes making them have support simultaneously at both ends, as it is apparent from Eq.~\rf{eq:edge1}. This is equivalent to the hybridisation of overlapping edge modes in local Hamiltonians, though now this effect cannot be removed by increasing the size $N$ of the system. Note that the special case with $J=\pm|\Delta|$ and $\mu=0$ gives rise to edge Majorana modes with $E_\pm=0$ and amplitudes $\psi^\pm_j$ that are ultra-localised at the ends of the chain, as expected from the Majorana description given in Fig.~\ref{fig:4site}~(Bottom).

\begin{figure}[t]
\centering
\includegraphics[width=0.85\columnwidth]{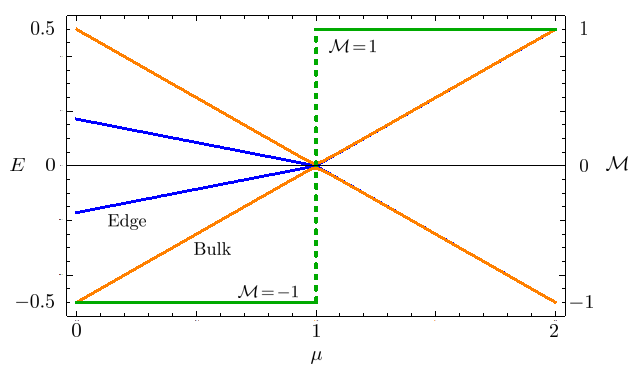}
\caption{The topological behaviour of the open chain as a function of the chemical potential $\mu$ for $J=1$, $\Delta=0.7$ and $N=100$. The system undergoes a topological phase transition at $\mu=J$. For $\mu<J$ the edge states (blue) are energetically distinct from the lowest bulk states (orange), while they coincide for $\mu>J$. The bulk gap closes as $N\to\infty$ at $\mu=J$. The sign, $\cal{M}$, of the Pfaffian (green) identifies the topological ${\cal M}=-1$ from the non-topological ${\cal M}=+1$ regimes of the open chain.}
\label{fig:winding}
\end{figure}

The presence of edge modes can also be diagnosed by a topological index characterising the phase of the system. Hamiltonian \rf{eq:powerlaw} belongs to symmetry class D in the classification of \citep{Schnyder,Kitaev2}. Both in $d=1$ and $d=0$, this symmetry class has a $\mathbb{Z}_2$ topological classification. In both cases the topological invariant is a parity, $\cal{M}$, that can be evaluated for non-degenerate open or closed chains in terms of the Pfaffian of the corresponding kernel Hamiltonian. Hence, it can be used directly for the case of the infinite-range Hamiltonian with $\alpha=0$.\cite{sym} 
By varying the chemical potential $\mu$ we observe that the parity becomes non-trivial, ${\cal M}=-1$, for $\mu<J$, as shown in Fig.~\ref{fig:winding}. In that regime the edge states, while they have non-zero energy, are energetically distinct from the bulk states. This is in contrast to the non-topological regime, ${\cal M}= +1$, where the edge states cease to exist. 

The topological phase of the non-local model is robust to weak disorder in the chemical potential. In this case the Majorana edge modes remain exponentially localised and the Majorana number, $\mathcal{M}$, remains non-trivial, as in the case with no disorder. 
%For strong disorder, greater than the order of the chemical potential, a phase transition occurs due to Anderson localisation and the topological character is lost.

%{\Kris The non-trivial ground states, and thus the non-zero energy Majorana modes, are robust to weak disorder. They are both exponentially localised and remain well separated from the gap, as in the case with no disorder. For strong disorder the effects of Anderson localisation are dominant, where ground states are no longer distinguishable from the bulk. This is evident in the Majorana number, $\mathcal{M}$, which has a non-trivial value and remains invariant only for weak disorder in the topological phase.}

%{\Kris Hence, we identify the topological phase of the system by the negative value of $Pf(A)$.}
%
%
%However, the long range parameter $\alpha$ can be continuously tuned from the infinite range $\alpha=0$ model to the local model, $\alpha\to\infty$, without the bulk gap closing whilst in a topological phase. The infinite range winding number is therefore defined as 
%\begin{align}
%\nu=\frac{1}{M}\oint_{BZ} \frac{S_y\partial S_z - S_z\partial S_y }{S_y^2+S_z^2}\,\, dp.
%\end{align}
%The case for non-zero $\alpha$ is considered next.

\begin{figure}[t]
\centering
\includegraphics[width=0.9245\columnwidth]{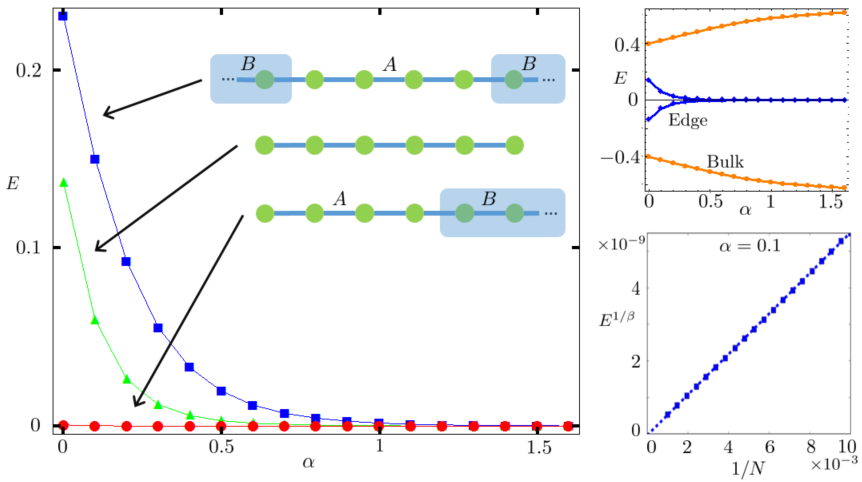}
\caption{With $J=1$, $\Delta=0.7$, $\mu=0.2$, and $N=400$. (Left) Energies of the Majorana edge modes for a chain with power law couplings parametrised by $\alpha$, corresponding to the physical spectrum (green), the entanglement spectrum with two cuts (blue) and the entanglement spectrum with one cut (red) for $A$ of size $N_A=204$. The relevant chains with their partitions for the entanglement spectra are also depicted. (Right Top) Ground state (blue) and first excited state (orange) for various $\alpha$. 
%The bulk gap remains intact for all $\alpha$ so there is no phase transition between the infinite range and local model. 
(Right Bottom) The edge state energy, linearised with $1/\beta$, as a function of system size for $\alpha = 0.1$, $\beta=0.137$. 
%For any non-zero power law coupling the edge state energies go to zero as the system size tends to infinity.
}
\label{fig:ealpha}
\end{figure}

{\bf \em Power law behaviour. --} 
We now show that the characteristics of the infinite-range model are robust when the couplings, in Eq.~\rf{eq:powerlaw}, take the more physical power law form. 
To demonstrate this we perform a numerical treatment of a finite chain for various $\alpha$. Even though a real spectrum of the semi-infinite chain cannot be obtained numerically, we show that, rather surprisingly, its properties are manifested in the entanglement spectrum~\citep{Li}. 
This spectrum is the set of eigenvalues of the entanglement Hamiltonian $H_E= - \ln \rho_A$, defined through the reduced density matrix $\rho_A=\tr_B\ket{\Psi}\!\bra{\Psi}$ for a chain bipartitioned into region $A$ and its complement $B$, where $\ket{\Psi}$ is the many-body ground state of the system~\cite{Titus}. Unlike the real spectrum of a finite chain, that has necessarily two ends, a choice of region $A$ can give rise to a single virtual boundary. 
%In this case the entanglement spectrum hosts a Majorana edge mode at zero entanglement energy for any value of $\alpha$, as shown in Fig.~\ref{fig:ealpha}. 
In this case the entanglement spectrum hosts degenerate zero modes, indicating the presence of Majorana edge modes in the same way as a local chain. Fig.~\ref{fig:ealpha} shows that this holds for any $\alpha$. Thus, the topological phase can be diagnosed with the same tools and criteria as a local chain.

The zero energy entanglement mode is in agreement with the analytical solution of the semi-infinite case with $\alpha=0$ derived in the previous section, thus demonstrating the robustness of the $E=0$ states to finite range couplings. 
It also demonstrates the physical relevance of the semi-infinite chain as its spectrum is manifested in the correlations of the ground state of a finite system. 
Interestingly, the entanglement eigenstate has a power-law decay, instead of the exponential decay which we obtained in Eq.~\rf{eq:semi}. 
This is a manifestation of the non-local couplings on the correlations across the partition (see Supplemental Material).
%This is attributed to the actual finite size of the chain (see Supplemental Material). 
For a bipartition in which the two regions interface at two separated points, the system is described by an open finite entanglement Hamiltonian. In this case the spectral degeneracy is lifted due to the long-range coupling between the edge states. 

The energy splitting of the Majorana edge states due to the infinite-range interactions remains present even if we employ decaying couplings with $\alpha\neq 0$. 
As $\alpha$ varies, there is a continuous interpolation between the finite energy $E_+$ given in~Eq.~\rf{eq:energy} at $\alpha=0$ and the zero energy at $\alpha\rightarrow \infty$. 
As one would expect from the argument of hybridisation, the energy will decay with $1/N^{\beta}$, where $\beta\approx\alpha$ and depends on the coupling parameters as well as $\alpha$. 
This property is shown in Fig.~\ref{fig:ealpha} for both the real and the entanglement spectra. 
Similar robustness in the values of $\alpha$ is exhibited by the topological index ${\cal M}$. 
As $\alpha$ increases, the topological index preserves its non-trivial value ${\cal M} =-1$ for all $\alpha$ when $\mu<J$. 
As can be seen in Fig.~\ref{fig:ealpha}~(Right Top), the limiting cases of $\alpha=0$ and $\alpha\to\infty$ are adiabatically connected in the topological sector.
%Both, the localised edge modes and the bulk gap, remain intact for all values of $\alpha$ so the system remains in the same topological phase, as shown in Fig.~\ref{fig:ealpha}~(Right Top). 

When $\alpha\neq 0$ the interaction between the end modes decay as their distance increases due to the power law dependence of the couplings. This causes the energy of the modes to tend to zero for increasing $N$, as shown in Fig.~\ref{fig:ealpha}~(Right Bottom). Nevertheless, for small enough $\alpha$ and $N$ the characteristic of non-zero energy will be present as the couplings will effectively have infinite range. These results remains qualitatively the same when the tunnelling couplings are assigned a shorter range interaction compared to the pairing ones.

{\bf \em Conclusions. --}
Our analytical treatment demonstrates that the infinite-range Kitaev chain is topological. It supports Majorana zero modes and it has a non-trivial topological index. The zero modes are exact in semi-infinite chains; a finite chain, regardless of its size, gives hybridised Majorana edge modes with non-zero energy. This result is derived exactly, presenting a theoretical framework for solving other long range problems. Importantly, the Majorana zero modes are identified in the entanglement spectrum of finite systems, for an appropriate partition. Thus, it presents a diagnostic tool for numerical simulations or experimental implementations of non-local models. The infinite range 1d Majorana chain is an archetype model against which other models can now be compared analytically.
%Hence, they can be employed to diagnose the topological phase in numerical simulations or physical implementations that necessarily deal with finite size systems. 

Notably, long-range interactions become increasingly relevant in current theoretical and experimental investigations. On the experimental front, Rydberg atoms, trapped ions, and atoms in optical cavities can be designed with interactions of arbitrary range. Many of these systems are engineered as spin chains, which can be mapped to long range fermionic chains~\cite{Porras,Islam}.
Alternatively, Shiba chains provide a solid-state system that naturally supports non-local interactions~\cite{Pientka1,Pientka2,Pientka3,Brydon,Rontynen}. Here, magnetic impurities placed on an s-wave superconductor have an effective Hamiltonian that contains long range hopping and pairing couplings.
On the theoretical front, we envision that the unique properties of non-local models can help us understand exotic effects of strongly correlated systems~\citep{Turner,Bernd}.

\begin{acknowledgments}
{\bf \em Acknowledgements. --} We would like to thank Hans Peter B\"uchler, Steve Simon, and Chris Turner for inspiring conversations. 
This work was supported in part by the EPSRC grant EP/I038683/1.
\end{acknowledgments}

\newpage

\section{A. Defining a generating function for non-local chains}
The non-local model can be understood analytically via a generating function method. The generating function provides a closed form equation containing information including the localisation and energy of eigenstates of the Hamiltonian. We begin with Hamiltonian (1), from the main text, on a finite chain of length $N$ and decay parameter $\alpha=0$ with $N\to\infty$ defining the semi-infinite chain. Due to particle-hole symmetry, the model can be represented in a spinor basis with 
\begin{align}
\chi_j=\begin{pmatrix}
a_j \\
a_j^{\dagger}
\end{pmatrix}\text{,}\quad \Gamma_1&=\begin{pmatrix}
\mu & 0 \\
0 & -\mu
\end{pmatrix}\quad\text{and}\quad \Gamma_2&=\begin{pmatrix}
J & -\Delta^* \\
\Delta & -J
\end{pmatrix}
\end{align}
In this new notation,
\be
H=\frac{1}{2}\sum_{j=1}^{N-1}\sum_{l=1}^{N-j}\left[\chi_j^{\dagger}\Gamma_2\chi_{j+l}+\chi_{j+l}^{\dagger}\Gamma_2^{\dagger}\chi_{j}\right]+\frac{1}{2}\sum_{j=1}^{N}\chi_{j}^{\dagger}\Gamma_1\chi_{j}
\ee
While $\Gamma_1$ and $\Gamma_2$ are not hermitian, the Hamiltonian is Hermitian. Here, the operators obey the usual fermion anti-commutator relation $\{\chi_i^{\alpha},\chi_j^{\beta}\}=\delta_{ij}\delta_{\alpha\beta}$ with $\chi_i^1=a_i$ and $\chi_i^2=a_i^{\dagger}$. Demand that $\ket{\psi}=\sum_m\chi_m^{\dagger}\psi_m\ket{0}$ is a ground state of $H$ with energy $E$ where $\psi_m=\begin{pmatrix}
\psi_m^A \\
\psi_m^B
\end{pmatrix}$ is a spinor containing particle and hole occupation amplitudes at site $m$. The eigenvalue equation $H\ket{\psi}=E\ket{\psi}$ along with the boundary condition $\psi_j=0$ for $j<1$ and $j>N$ gives the recurrence relation 
\be
0=\sum_{l=1}^{N}\Big[\Gamma_2\psi_{j+l}+\Gamma_2^{\dagger}\psi_{j-l}\Big]+(\Gamma_1-2E)\psi_j
\ee
which is valid for $1\leq j \leq N$. As the infinite-range model in a fermionic basis is equivalent to the fully connected graph each term requires information from every other term to be determined. Simplify this with the substitution $\Psi_j=\sum_{k=j}^{N}\psi_k$ to give
\be
0=M\Psi_{j}+K\Psi_{j+1}+\Gamma_2^{\dagger}\Psi_1
\ee
where $M=\Gamma_1-\Gamma_2^{\dagger}-2E$ and $K=\Gamma_2-\Gamma_1+2E$. 

From a recurrence relation we can define a generating function. Multiply the recurrence by $z^j$ for $z\in\mathbb{C}$ and sum over all $j$
\begin{align}
0&=\sum_{j=1}^N\big[z^jK\Psi_{j+1}+zz^{j-1}M\Psi_{j}+z^j\Gamma_2^{\dagger}\Psi_1\big]\nonumber\\
0&=\sum_{j=1}^N\big[z^{j-1}K\Psi_{j}+zz^{j-1}M\Psi_{j}+z^j\Gamma_2^{\dagger}\Psi_1\big]-K\Psi_{1}\nonumber\\
0&=(K+zM)G(z)+\left[-K+\Gamma_2^{\dagger}\frac{z(1-z^N)}{1-z})\right]\Psi_1,
\end{align}
with the generating function defined as 
\begin{align}
G(z)&=\sum\limits_{j=1}^Nz^{j-1}\Psi_j\\
&=(K+zM)^{-1}\left[K-\Gamma_2^{\dagger}\frac{z(1-z^N)}{1-z})\right]\Psi_1.
\end{align}

Expand $A=K+zM$, find its inverse and let $\Psi_1=\begin{pmatrix}
\phi_1 \\
\phi_2
\end{pmatrix}$ to give
\begin{align}\label{eq:fullGF}
G&(z)=\frac{1}{\det A}\left[\begin{matrix}
\Sigma(1-z) & \Delta^*(1+z)\\
-\Delta(1+z) & -\bar{\Sigma}(1-z)
\end{matrix}\right]\nonumber\\
&\times\left[\begin{pmatrix}
-\bar{\Sigma}\phi_1-\Delta^*\phi_2\\
\Delta\phi_1+\Sigma\phi_2
\end{pmatrix}-\frac{z(1-z^N)}{1-z}\begin{pmatrix}
J\phi_1+\Delta^*\phi_2\\
-\Delta\phi_1-J\phi_2
\end{pmatrix} \right],
\end{align}
where $\det A=[4E^2-(J-\mu)^2](1-z)^2+|\Delta|^2(1+z)^2$,  $\Sigma=-J+\mu+2E$ and $\bar{\Sigma}=-J+\mu-2E$.
This generating function represents both a finite chain and a semi-infinite chain where the latter is recovered upon setting $N\to\infty$.

\section{B. Edge Modes on a Non-Local Semi-Infinite Chain}
To determine the presence of edge modes we will study the singular behaviour of the generating function corresponding to the non-local model defined in Sec. A. The conditions for a semi-infinite non-local chain to host edge modes are identical to those given for a local chain in \citep{Pershoguba}. However, first they must be shown to be equivalent. 

The generating function for a non-local semi-infinite chain with $N\to\infty$ is defined as $G(z)=\sum\limits_{j=1}^{\infty}z^{j-1}\Psi_j$ where $\Psi_j=\sum_{k=j}^{\infty}\psi_k$ is the sum of the single site fermionic spinors. For a mode exponentially localised at the $j=1$ boundary use the ansatz $\psi_j=\frac{1}{z_1^j}$ with $|z_1|>1$. Together with some algebra the summed state can be rewritten as
\begin{align}
\sum_{k=j}^{\infty}\psi_k&=\sum_{k=j}^{\infty}\frac{1}{z_1^k}\nonumber\\
&=\frac{1}{z_1^j}\sum_{l=0}^{\infty}\frac{1}{z_1^l}, \Sp \text{for } l=k-j, \nonumber\\
&=\frac{1}{z_1^j}\frac{z_1}{z_1-1},
\end{align}
which on returning to the generating function gives 
\begin{align}
G(z)&=\sum_{j=1}^{\infty}\frac{z^{j-1}}{z_1^j}\frac{z_1}{z_1-1}\nonumber\\
&=\frac{z_1}{z_1-1}\sum_{j=1}^{\infty}z^{j-1}\psi_j,
\end{align}
where $g(z)=\sum_{j=1}^{\infty}z^{j-1}\psi_j$ is the definition of a generating function for a local chain. We now turn to the local generating function $g(z)$. 

{\bf Proposition:}
A rational generating function, $g(z)$, corresponds to an edge mode, $\psi_j\xrightarrow{\scriptscriptstyle j\to\infty}0$, if and only if all the poles, $z_{i}$, of $g(z)$ have absolute values greater than one, $|z_i|>1$ $\forall$ $i$.

{\bf Proof:}
Any rational function with poles, $z_i$, can be transformed to the form $g(z)=\sum_i\frac{f_i(z)}{(z-z_i)^{n_i}}$ where $f_i(z)$ is a polynomial in $z$ and $n_i$ is the order of the pole $q_i$.
Consider a generating function with a first order pole only. It follows that
\begin{align}
g(z)&=\frac{z_1}{z_1-z}\\
&=\frac{1}{1-\frac{z}{z_1}}\\
&=\sum_j\left(\frac{z}{z_1}\right)^j.
\end{align}
From the definition of the local generating function we have that $\psi_j=\frac{1}{z_1^{j-1}}$. Therefore, either
\begin{align}
&|z_1|<1 \qquad \implies \qquad \psi_j\xrightarrow{\scriptscriptstyle j\to\infty}\infty,\\
&|z_1|=1 \qquad \implies \qquad \psi_j=e^{ikj},\\
&|z_1|>1 \qquad \implies \qquad \psi_j\xrightarrow{\scriptscriptstyle j\to\infty}0,
\end{align}
corresponding to a diverging (or converging on the opposite boundary), a bulk or an edge mode solution respectively.
All other generating functions can be reduced to this by partial fraction decomposition so this completes the proof.

It now remains to analyse the singular points (or poles) of the non-local generating function to understand properties of a chain with an edge mode.  Solving $\det A=0$ gives the poles $z_1$ and $z_2$ of $G(z)$ but Vieta's formula gives $z_1z_2=1$ so either $|z_1|=|z_2|=1$ or $|z_1|<1<|z_2|$ or $|z_2|<1<|z_1|$. For the existence of a mode localised to a single edge there must be a pole for $|z_i|>1$ only. Therefore demand that the $|z|\leq1$ mode is cancelled by the numerator of $G(z)$.

As we are interested in cancelling the $|z|<1$ poles we restrict to that domain for which $z^N\to0$. Of course, with $N\to\infty$ there exists singular points for all $|z|>1$, however this does not affect the existence of edge modes as they satisfy the condition found at the beginning of this Appendix. Expanding the generating function in full gives 
\begin{align}
G(z)=\frac{1}{(1-z)\det A}\left[(1-z)P(z)-zQ(z)\right]
\end{align}
with $P(z)$ and $Q(z)$ both having components linear in $z$,
\begin{align*}
P(z)&=\left[\begin{smallmatrix}
-\Sigma(\bar{\Sigma}\phi_1+\Delta^*\phi_2)(1-z)+\Delta^*(\Delta\phi_1+\Sigma\phi_2)(1+z)\\
\Delta(\bar{\Sigma}\phi_1+\Delta^*\phi_2)(1+z)-\bar{\Sigma}(\Delta\phi_1+\Sigma\phi_2)(1-z)
\end{smallmatrix}\right]\\
Q(z)&=\left[\begin{smallmatrix}
\Sigma(J\phi_1+\Delta^*\phi_2)(1-z)-\Delta^*(\Delta\phi_1+J\phi_2)(1+z)\\
-\Delta(J\phi_1+\Delta^*\phi_2)(1+z)+\bar{\Sigma}(\Delta\phi_1+J\phi_2)(1-z)
\end{smallmatrix}\right].
\end{align*}
In order to cancel the $|z|<1$ singular point and the $z=1$ pole the rows of the quadratic $(1-z)P(z)+zQ(z)$ must be proportional. Set proportionality relations between coefficients of $1$, $z$ and $z^2$ and eliminate the constant of proportionality to solve for the energy. We find that the only energy solution is $E=0$.

As the generating function is proportional to $\Psi_1$ all $\Psi_i$ are proportional to $\Psi_1$. Setting $\phi_2=1$ along with the zero energy solution gives $\phi_1=\pm e^{-i\theta}$ where $\theta$ is the complex phase of the superconducting gap $\Delta=|\Delta|e^{i\theta}$. The spinor for a single site can then be found as the difference of consecutive sums
\begin{align}
{\psi}_j=\Psi_j-\Psi_{j+1}\sim\begin{pmatrix}
\pm e^{-i\theta}\\
1
\end{pmatrix}
\end{align} 
Earlier we found that a mode localised at $j=1$ decays into the bulk with the remaining $|z_1|>1$ pole which corresponds to the positive $\phi_1$ solution. If $|z_1|<1$ the mode will localise at a $j=N$ boundary of a semi-infinite chain with the negative $\phi_1$ solution. Therefore, a zero energy mode localising at $j=1$ has a spinor of the form 
\begin{align}
\psi_j=z_1^{-j}\begin{pmatrix}
e^{-i\theta}\\
1
\end{pmatrix},
\end{align}
up to normalisation. Demanding that the remaining singular point, $z_1$, has absolute value greater than $1$ gives the condition for an edge mode
\begin{align}
\mu<J.
\end{align}

\section{C. Symmetries of a Finite Non-Local Chain}
The Hamiltonian for a non-local finite chain has certain symmetries that give specific conditions on eigenstates that are not present on a semi-infinite chain. We consider two sets of transformations of an open chain with boundaries at $j=1$ and $j=N$. Let the first set of transformations be represented by $T$ that are described by
\begin{align}
j&\to N-j+1\nonumber\\
l&\to -l\nonumber\\
a_j&\to -a_j\nonumber\\
a_j^{\dagger}&\to a_j^{\dagger}.
\end{align}
Acting $T$ on the Hamiltonian gives
\begin{align}
H_T=T^{\dagger}HT=-H,
\end{align}
so the operator satisfies $\lbrace H,T\rbrace=0$. An eigenstate of the Hamiltonian $\ket{\psi}=\sum_j(a_j^{\dagger} a_j)\begin{pmatrix}
\psi_j^A\\
\psi_j^B
\end{pmatrix}\ket{0}$ with its corresponding energy $E$ leads to $\ket{\psi}=-\ket{\psi}_T$ where $\ket{\psi}_T=T\ket{\psi}$ is an eigenstate of $H_T$. The equality gives information about particle and hole amplitudes 
\begin{align}
\psi_j^A&=-\psi_{N-j+1}^A\\
\psi_j^B&=\psi_{N-j+1}^B
\end{align}
From the first condition above it is then clear that one component of $\Psi_1$ is zero,
$
\sum\limits_{j=1}^N\psi_j^A=0,
$
with the other component non-zero. This symmetry and therefore this condition is unique to the finite chain and does not hold for a semi-infinite chain.

Instead the Hamiltonian can be transformed with 
\begin{align}
j&\to N-j+1\nonumber\\
l&\to -l\nonumber\\
a_j&\to a_j\nonumber\\
a_j^{\dagger}&\to -a_j^{\dagger}.
\end{align}
This transformation also results in $-H$ and gives the condition $\sum\limits_{j=1}^N\psi_j^B=0$, with the top component non-zero.

\section{D. Divergences of the Finite Non-Local Chain}
As the generating function for a finite chain is defined differently to the semi-infinite chain the singular points may behave differently. Here we study the divergences of a generating function defined by $G(z)=\sum_j^Nz^{j-1}\Psi_j$ with 
\begin{align}
\Psi_j&=\sum_{l=j}^{N}\psi_{l}\nonumber\\
&=\sum_{l=1}^N\psi_l-\sum_{l=1}^j\psi_l+\psi_j.
\end{align}  
Thus, the generating function can be split up into terms
\begin{align}
G(z)=\sum_{j=1}^Nz^{j-1}\sum_{l=1}^N\psi_l-\sum_j^N\sum_{l=1}^jz^{j-1}\psi_l+\sum_j^Nz^{j-1}\psi_j.
\end{align}
Due to non-local coupling between edges and symmetries of the chain we find that a single mode must be localised at both boundaries with equal amplitudes so use the ansatz $\psi_j=z_1^{-l}+z_1^{l-N-1}$ with $|z_1|>1$ and $\bar{z}_1=\frac{1}{z_1}<1$. This is verified by numerical studies so is a good choice of ansatz. With this the generating function becomes
\begin{align}
G(z)=\frac{1-z^N}{1-z}\frac{\bar{z}_1^N-1}{1-z_1}+\frac{\bar{z}_1^N}{1-z_1}\frac{1-(zz_1)^N}{1-zz_1}\\+\frac{1}{z_1-1}\frac{1-(z\bar{z}_1)^N}{1-z\bar{z}_1},
\end{align}
and $$\lim_{N\to\infty}\bar{z}_1^N\to 0$$ so
\begin{align}
G(z)=\frac{1-z^N}{1-z}\frac{1}{z_1-1}+\frac{1}{z_1-1}\frac{1-(z\bar{z}_1)^N}{1-z\bar{z}_1}.
\end{align}
It can be seen immediately that the generating function has no pole for any $|z|<1$, in particular there is no pole at $\bar{z}_1$. Studying the divergences further gives that as
\begin{align}
z\to1 && G&\to\frac{N}{z_1-1}+\frac{1}{(z_1-1)(1-\bar{z}_1)},\\
z\to z_1 && G&\to\frac{z_1^N-1}{(1-z_1)^2}+\frac{N}{z_1-1}.
\end{align} 
Thus, the $z=z_1$ pole diverges exponentially with $N$ and the $z=1$ pole diverges polynomially with $N$. For $z_1=1$ the generating function has poles for all $z$ and is a critical point. 

\section{E. Energy and Localisation for Finite Chains}
The generating function was found in Appendix A and defined as a sequence of summed states $G(z)=\sum\limits_{j=1}^Nz^{j-1}\Psi_j$, where the sum is decreasing in number of states present $\Psi_j=\sum\limits_{l=j}^N\psi_l$. The sum of all onsite spinors can be simplified by $\Psi_1=\begin{pmatrix}
\phi_1\\
\phi_2
\end{pmatrix}$ so that the generating function takes the same form as Eq.~\eqref{eq:fullGF}
\begin{align}
G&(z)=\frac{1}{\det A}\left[\begin{matrix}
\Sigma(1-z) & \Delta^*(1+z)\\
-\Delta(1+z) & -\bar{\Sigma}(1-z)
\end{matrix}\right]\nonumber\\
&\times\left[\begin{pmatrix}
-\bar{\Sigma}\phi_1-\Delta^*\phi_2\\
\Delta\phi_1+\Sigma\phi_2
\end{pmatrix}-\frac{z(1-z^N)}{1-z}\begin{pmatrix}
J\phi_1+\Delta^*\phi_2\\
-\Delta\phi_1-J\phi_2
\end{pmatrix} \right],
\end{align}
with $\Sigma=(-J+\mu+2E)$ and $\bar{\Sigma}=(-J+\mu-2E)$.

In Appendix C and D we derived conditions required for the existence of edge modes. By studying the symmetries of the Hamiltonian we found that either $\phi_1=0$ or $\phi_2=0$ gives the general solutions of a finite chain. After studying the divergences of the generating function there must exist a linearly diverging pole at $z=1$ and an exponentially diverging singular point at $|z_1|>1$ for an edge mode to exist. Isolate the pole that we wish to cancel, so restrict ourselves to $|z|<1$. This sets $z^N\to0$ for large $N$ whilst retaining the $z=1$ pole. Setting one of the conditions found in Appendix C, $\phi_2=0$, on the generating function gives
\begin{align}
G_1(z)=X\left[\begin{smallmatrix}
z^2(\Sigma J-\Sigma\bar{\Sigma})+z(2\Sigma\bar{\Sigma}+|\Delta|^2-\Sigma J)+(|\Delta|^2-\Sigma\bar{\Sigma})\\
z^2(\Delta J-\Delta\bar{\Sigma})+z(\Delta J+\Delta\bar{\Sigma})
\end{smallmatrix}\right],
\end{align}
where $X=\frac{\phi_1}{(1-z)\det A}$. The bottom row has zeroes for $z=0$ and $z=\frac{\bar{\Sigma}+J}{\bar{\Sigma}-J}$. The zeroes of the rows must be the same and equal the pole to cancel it. Begin with $z=0$ and set it as a zero of the top row to give the energy $E=\pm\frac{1}{2}\sqrt{(\mu-J)^2-|\Delta|^2}$.
This is the energy of a mode with $z=0$ a common zero. However cancelling this with a pole at $z=0$ leaves $|z_1|>1$, $|z_2|<1$ and $z_3=1$ poles of the generating function. For an edge mode the generating function only requires poles for $|z_i|>1$ and at $z_i=1$ so $z=0$ should not be a zero of both rows. 

Making the $z\neq 0$ zero of the bottom row a zero of the top row gives the energy solution
\begin{align}
E_+=\frac{J-\mu}{2}\hspace{2mm}\frac{J^2-|\Delta|^2}{J^2+|\Delta|^2}.
\end{align}

Equally we could have set the condition $\phi_1=0$ from Appendix C giving the negative energy solution
\begin{align}
E_-=-\frac{J-\mu}{2}\hspace{2mm}\frac{J^2-|\Delta|^2}{J^2+|\Delta|^2}
\end{align}
Demanding that the non-zero $z$ solution from the top row of $G_1(z)$ has an inverse that is a pole with $|z|>1$ gives the condition for an edge mode
\begin{align}
\mu<J.
\end{align}
This condition persists for both energy solutions.

From this, it remains to find the form of the on site spinors to give localisation. As the generating function is proportional to $\Psi_1$ then all $\psi_j$ are too as $\psi_j=\Psi_{j}-\Psi_{j+1}$. With the symmetries of the chain and the condition for one component of $\Psi_j$ being $0$ from the previous appendices, the modes take the form $\psi_j\propto\begin{pmatrix}
1\\
a
\end{pmatrix}$ at one end of the chain and $\psi_{N-j+1}\propto\begin{pmatrix}
1\\
-a
\end{pmatrix}$ at the other end. Here, the free component has been set to $1$ on a positive energy chain.

The generating function is proportional to $\Psi_1$ for all $z$. We can therefore take any value of $z$ that simplifies the generating function and find what form $\psi_j$ will take. The function is simplified greatly at the linearly diverging $z=1$ pole so consider the limit of $G(z)$ at this point with $\phi_2=0$,
\begin{align}
\lim_{z\to1}G_1(z)&=\frac{\phi_1}{\det A}\frac{2}{\Delta}\left[\begin{pmatrix}
\Delta^*\\
\bar{\Sigma}
\end{pmatrix}-N\begin{pmatrix}
-\Delta^*\\
- J
\end{pmatrix} \right].
\end{align}
In this limit the dominant term of the generating function is the one with the pole. The generating function can be simplified further to
\begin{align}
\lim_{z\to1}G(z)&=C N\begin{pmatrix}
1 \\
\frac{J}{\Delta^*}
\end{pmatrix}
\end{align}
where $C$ contains the constant terms appearing after normalisation. 
As each site spinor is proportional to the generating function, modes must be of the form $\psi_j\sim\begin{pmatrix}
1\\
\frac{J}{\Delta^*}\end{pmatrix}$. From the analysis of symmetries and divergences we know that the edge modes decay with the remaining pole that satisfies $|z_1|>1$ so the final form of a spinor with $E_+$ is
\begin{align}
\psi^{+}_j=z_1^{-j}\begin{pmatrix}
1\\
\frac{J}{\Delta^*}
\end{pmatrix}+ z_1^{j-N-1}\begin{pmatrix}
1\\
-\frac{J}{\Delta^*}
\end{pmatrix}
\end{align}
One can also take $\phi_1=0$ to find an edge mode with energy $E_-$ and its spinor of the form
\begin{align}
\psi^{-}_j=z_1^{-j}\begin{pmatrix}
\frac{J}{\Delta}\\
1
\end{pmatrix}+ z_1^{j-N-1}\begin{pmatrix}
-\frac{J}{\Delta}\\
1
\end{pmatrix}
\end{align}
where in both cases $z_1=\frac{\bar{\Sigma}-J}{\bar{\Sigma}+J}$. These solutions agree exactly with numerical results.

\begin{figure}[t]
\centering
\includegraphics[width=0.8\columnwidth]{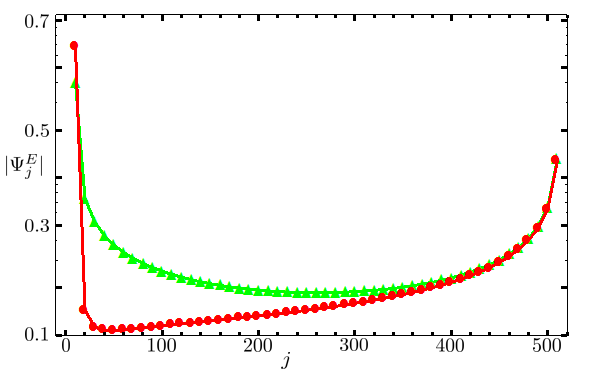}
\caption{Ground state eigenvector of the entanglement Hamiltonian $H_E$ with $N=1000$, $J=1$, $\Delta=0.7$ and $\mu=0.2$. The region A has 504 sites with (red circles) a single cut forming a single virtual edge (green triangles) two cuts forming two virtual edges. In both cases the edge mode decays polynomially and gives a degenerate ground state in the case of a single cut and a non-degenerate ground state for two cuts. This can be attributed to a coupling between virtual edges causing the spectrum to split.}
\label{fig:power}
\end{figure}
\section{F. Power-law Entanglement Eigenstates of Finite Non-Local Chains}

Finally, we study the entanglement ground state and compare its localisation properties with the physical ground state. The entanglement Hamiltonian $H_E= - \ln \rho_A$ is defined via the reduced density matrix $\rho_A=\tr_B\ket{\Psi}\!\bra{\Psi}$ for a chain bipartitioned into region $A$ and its complement $B$, where $\ket{\Psi}$ is the many-body state of the system. The ground state eigenvector is the sate corresponding to the lowest energy eigenvalue of the Hamiltonian $H_E$. Including non-local couplings to the local Kitaev chain results in a degenerate entanglement spectrum ground state that points to topologically protected edge modes. The localized eigenstates of the entanglement Hamiltonian, however, decay polynomially even if the edge eigenstates of the actual Hamiltonian are exponentially localized. Correspondence between the physical and entanglement eigenstates breaks down due to the long range nature of the model. There are currently no known gapped local models with this property making it a signature of non-locality. In Fig.~\ref{fig:power} we show the ground state for different entanglement cuts exposing the polynomial nature at both physical and virtual ends of the partition.


\begin{thebibliography}{43}
\bibitem[]{sym} Note that it is possible to assign a $\mathbb{Z}$ classification to the model by defining a local chiral symmetry. For a short range chain the model would then belong to the BDI symmetry class. However, terms may be added to the Hamiltonian that break this chiral symmetry, so that the chain remains in class D. This $\mathbb{Z}_2$ symmetry sufficiently describes the topological phase of the chain and can be accurately probed by the Pfaffian invariant.
\end{thebibliography}
\end{document}